\documentclass[12pt]{article}
\usepackage{graphicx,floatflt,amssymb,epsf,psfig,rotate} 
\def\ovl{\overline}
\def\be{\begin{equation}}
\def\ee{\end{equation}}
\def\ba{\begin{eqnarray}}
\def\ea{\end{eqnarray}}

\def\al{\alpha}
\def\Z{M_Z}

\def\br{\begin{array}}
\def\er{\end{array}}
\def\bc{\begin{center}}
\def\ec{\end{center}}

\def\ps{{\cal G}_{PS}}
\def\lr{{\cal G}_{LR}}
\def\std{{\cal G}_{std}}
\def\DESepsf(#1 width #2){\epsfxsize=#2 \epsfbox{#1}}
\begin{document}
%\thispagestyle{empty}
%\begin{flushright} 
%\texttt{hep-ph/yymmnnn}\\ 
%CU-PHYSICS-20-2006\\ 
%HRI-P-06-12-001 \\
%\end{flushright}  
%\vskip 30pt 
\begin{center}
{\Large \bf Inverse see-saw, leptogenesis,~
observable proton decay and $\Delta^{\pm\pm}_{\rm R}$ 
 in SUSY  $SO(10)$ with heavy $W_R$} 
\vskip .25in
\bf{\bf  Mina K. Parida${}^1$,   Amitava 
Raychaudhuri${}^{2,3}$\\}
\vskip .15in
{\sl ${}^1$ National Institute of Science Education and Research,\\
Institute of Physics Campus, Sachivalaya Marg, Bhubaneswar 751005,  India}\\
{\sl ${}^2$ Harish-Chandra Research Institute, Jhunsi, Allahabad
211 019, India}\\
{\sl ${}^3$ Department of Physics, University of Calcutta,
Kolkata 700 009, India}\\
\end{center}
\vskip .1in

\begin{abstract}
\noindent
   We explore the prospects of low-scale leptogenesis in a class
of supersymmetric $SO(10)$  models using extra  singlet neutrinos
($T_i, i=1,2,3$) and the Higgs representations ${\bf
{126}_H\oplus}$ ${\bf {\overline {126}}_H}$ as well as ${\bf {16}_H\oplus
{\overline {16}}_H}$. A singlet neutrino,
which we show can be as light as 10$^5$-$10^6$ GeV, decays 
through its small mixings with right-handed (RH)
neutrinos creating a lepton asymmetry which is explicitly shown to be
flavor dependent. While the doublet vacuum expectation value
({\em vev}) in ${\bf {\ovl {16}}_H}$ triggers the  generation of
desired mixings, it also induces a large RH triplet {\em vev}  that
breaks the left-right intermediate gauge symmetry and gives large
right-handed  neutrino masses. Manifest unification of gauge
couplings and generation of heavy RH neutrino masses are achieved
by purely renormalizable interactions.  The canonical (Type-I)
see-saw contributions to the light neutrino mass matrix 
cancel out while the  Type-II see-saw contribution is
negligible. Determining the parameters of the dominant inverse
see-saw formula by using the underlying quark-lepton symmetry and
neutrino oscillation data, we show how leptogenesis under the
gravitino constraint is successfully implemented. New
formulas for the decay rate and the asymmetry parameter are
derived  leading to baryon asymmetry within the observed range
without invoking a resonant condition on RH  neutrinos.  The
model is found to work for hierarchical as well as inverted
hierarchical light neutrino masses.  Testable predictions of the
model are RH  doubly charged Higgs bosons which may be
leptophilic and  accessible to the Tevatron, LHC or a linear
collider. In a model-independent manner, the Drell-Yan pair
production cross section  at Tevatron or LHC is shown to be
bounded between $59\%$-$79\%$ of their left-handed counterparts
with same mass.~In contrast to single-step breaking SUSY GUTs,
which predict a long proton lifetime for the decay $p\to
e^+\pi^0$,  here this lifetime is substantially reduced, bringing
it within one order of the current experimental limit.\\

\vskip 10pt \noindent 
\texttt{PACS Nos:~ 14.60.Pq, 12.10.Kt, 13.35.Hb } \\ 
\texttt{Key Words:~~ Neutrino mass, Leptogenesis, Grand Unified
Theories}
\end{abstract}
\newpage

\renewcommand{\thesection}{\Roman{section}}
\section{Introduction} 

${SO(10)}$ \cite{gfm} with supersymmetry (SUSY)  has been at the
centre of attention for a number of  attractive feaures.  It
contains just one right-handed (RH) neutrino per generation in
its spinorial representaion $\bf {16}$. With Pati-Salam
\cite{ps} and  left-right gauge symmetries \cite{lr} as its
subgroups, in addition to unification of the three forces of
Nature, ~it predicts high scale unification of quark and lepton
masses ~\cite{babupati} and has the potential to explain the
origin of parity ($\equiv {\rm P}$) and CP violations.  Using the
Higgs representations $\bf {{126}_H \oplus \ovl {126}_H}$ and
${\bf {10}_H}$, it reproduces the small masses and large mixings
of neutrinos through Type-I and Type-II see-saw mechanisms and
their extensions \cite{type1, type2, bajc, goh}.  It has been
also shown that all the fermion masses can be fitted through SUSY
$SO(10)$ by using suitable Higgs representations \cite{goh}.
Another interesting aspect of the theory is that the observed
tiny amount of matter-antimatter asymmetry of the universe can be
naturally explained through leptogenesis \cite{fuku-yana} and
sphaleron effects \cite{pati, ji, barr}.

In these theories neutrino masses indicated by  oscillation data
require the canonical see-saw scale of right-handed neutrinos to
be in the range of $M_R \sim 10^{13}$ - $10^{15}$ GeV. This also
sets the scale for the masses of associated Higgs triplets
carrying  $B-L= \pm 2$.  This scale of neutrino mass generation
is high in models with variants of the canonical see-saw
\cite{bajc,goh,barr} as well.

It is well known that the scale of leptogenesis through
right-handed  neutrino decays and canonical see-saw is
constrained from below leading to the lower bound on the lightest
RH neutrino mass $M_{N_1} \ge 10^9$ GeV \cite{di}. This in turn
requires the reheating temperature of the universe after
inflation to be at least $T_{RH} \sim 10^9$ GeV.  On the other
hand, big-bang nucleosynthesis in SUSY theories sets a severe
constraint on the gravitino mass and the reheating temperature
leading to the upper bound $ T_{RH}\le 10^7$ GeV
\cite{khlopov}.  While thermal leptogenesis in  SUSY $SO(10)$
with high see-saw scale  easily satisfies the lower bound, the
tension with  the gravitino constraint is manifest.

Independent of quark-lepton unified theories, the question of
baryogenesis via leptogenesis has been addressed in the context
of the Standard Model (SM) and the Minimal Supersymmetric
Standard Model (MSSM) \cite{buch} where freedom in the choice of
Dirac neutrino Yukawa couplings permits fine-tuning them
to very small values. In most of these models TeV scale resonant
leptogenesis \cite{reso} is realized by  degeneracy between
right-handed neutrino masses.  A major difficulty in having low-scale 
leptogenesis
in SUSY $SO(10)$ is the absence of such freedom because the
underlying quark-lepton symmetry requires these Yukawa couplings
to be of the same order as  the corresponding up-quark Yukawa
couplings.  This latter difficulty persists even in some
non-canonical see-saw models and several attempts have been made
to bring down the scale of leptogenesis \cite{valle}.

Another difficulty in renormalizable SUSY $SO(10)$ arises from
the gauge coupling unification constraint and the need for an
$SU(2)_R\times U(1)_{B-L}$ breaking intermediate scale that
generates RH neutrino masses through renormalizable Majorana type
interactions. It has been found that manifest  unification of
gauge couplings is spoiled in the presence of Higgs triplets of
${\bf {126}_H \oplus \overline {126}_H}$ with intermediate
symmetries such as  $SU(2)_L\times SU(2)_R \times  U(1)_{B-L} \times
SU(3)_{3C}$ or $SU(2)_L\times SU(2)_R \times SU(4)_{4C}$ even at
scales a few orders lower than the GUT-scale unless, in the first
case, the LR gauge theory and $SO(10)$ are extended to include
$S_4$ flavor symmetry \cite{parida} or additional light scalar
degrees of freedom are introduced at lower scales \cite{lee, mprs}.  On
the other hand, there are  a number of models with light
right-handed gauge bosons \cite{malinsky, dev} in which in place
of the Higgs triplets with $B-L = \pm 2$ there are doublets
carrying $B-L = \pm 1$.  In contrast to the above scenarios, here
we are interested in SUSY $SO(10)$ with both doublet and triplet
scalars.

The Higgs triplets in  ${\bf {126}_H}$ and  ${\bf \ovl {126}_H}$
representations include doubly charged bosons,
$\Delta^{\pm\pm}$. Dedicated searches for such doubly charged
scalars are being carried out at the Fermilab Tevatron
\cite{Tevatron}. Both the statistics and the energy reach are
expected to be further enhanced at the CERN LHC. However, the
high see-saw scale SUSY $SO(10)$ models will offer no prospects
for these searches as the corresponding masses are large,
$M_{\Delta} \ge 10^{11}$ GeV, while in the class of
low intermediate scale SUSY $SO(10)$ models where only RH
doublets in ${\bf {16}_H\oplus {\ovl {16}_H}}$ are used near the
TeV scale \cite {malinsky, dev, malinsky2} no doubly charged
Higgs bosons are present.

In this paper we address the issues of neutrino masses and
mixings, low-scale leptogenesis consistent with the gravitino
constraint, manifest unification of gauge couplings through
renormalizable interactions, and testable experimental signatures
of the proposed model at the Tevatron, LHC or ILC.  We construct
the desired SUSY $SO(10)$ model including the RH triplets in
${\bf 126_H \oplus \ovl {126}_H}$ as well as the RH doublets in
${\bf 16_H \oplus \ovl {16}_H}$, and  three singlet fermions
($T_i, ~i=1,2,3$) \cite{rnm3}.  We find that  a singlet fermion
in the mass range $M_T=10^5$ - $10^6$ GeV can go out of
equilibrium to generate lepton asymmetry; its decay is naturally
suppressed by  small   mixing with heavy right-handed neutrinos
($N_i$).  The vacuum expectation value  of the RH-doublet in
${\bf {\ovl {16}}_H}$ (or ${\bf {16}_H}$) responsible for this
desired small mixing also induces a large {\em vev} of the
RH-triplets in ${\bf {126}_H}$ (or ${\bf {\ovl {126}}_H}$).  This
breaks $SU(2)_R\times U(1)_{B-L} \to U(1)_Y$ leading to large
RH-neutrino masses through renormalizable interactions. We find
that although  heavy right-handed neutrinos are present in the
model, the Type-I see-saw contributions to the neutrino mass
cancel out as has been observed in the context of the Standard
Model or its extension \cite{ellis, kang}. The Type-II
contribution is also found to be negligible. The dominant
contribution to light neutrino masses arises through an inverse
see-saw  which has attracted considerable attention over the
recent years \cite {dev, lindner,deppisch}. 

 In an earlier work by us and S. K. Majee  it was found that
gauge coupling unification  with threshold-like behavior would be
possible through the presence of two non-renormalizable dim.-5
operators \cite{mpr}. Here, without using any dim.5 operators,
we obtain manifest  unification of gauge couplings  in the
renormalizable theory with asymmetric left-right intermediate
gauge symmetry ($g_{2L} \neq g_{2R}$) operative at any scale
between $10^9$ and $10^{15}$ GeV. Further, while lepton asymmetry
was computed through solutions of Boltzmann equation \cite{mpr} with
an assumption about the asymmetry parameter, in this
work we derive new analytic formulas for the decay rate and the
CP-asymmetry parameter and find that they are explicitly  flavor
dependent. We then show  analytically that when the model
parameters estimated using the neutrino oscillation data are used
in our new formula, the model yields desired values of the
CP-asymmetry parameter leading to the observed baryon to photon
density ratio.  In addition, we demonstrate that the model is
consistently successful for both hierarchical as well as
invertedly hierarchical light neutrino masses.  The model leaves
its testable signature at the LHC, Tevatron and ILC \cite
{Tevatron, LHC, ILC} through doubly charged right-handed Higgs
scalars $\Delta_R^{\pm\pm}$ in the mass range of $100$ GeV to a
few TeV.  Since the decay mode $\Delta_R^{\pm\pm} \to W^{\pm}_R
W^{\pm}_R$ is kinematically forbidden these Higgs bosons are
leptophilic and predominantly result in like-sign charged
bilepton pairs $\Delta_R^{\pm\pm} \to l_R^{\pm}l_R^{\pm}$. The
absence of light $\Delta_L^{\pm\pm}, \Delta_L^{\pm}$  states and
also the absence of left-handed bilepton pairs in the decays
would provide signatures specific to this model which are
different from other bilepton production modes.

In a model-independent manner without using any structure
function data, we show analytically that the Drell-Yan hadronic
pair production cross section for these RH Higgs bosons is
bounded between $59\%-79\%$ of that for a left-handed boson of
similar mass.

It is found that, triggered by low mass RH doubly charged Higgs,
at the unification scale the GUT coupling  lies in the strong but
perturbative regime and the gauge-boson mediated proton decay
rate is enhanced. The lifetime $\tau_p(p\to e^+\pi^0)$ is shorter
and remains within one order of the current experimental limit;
this can be reached by the ongoing or planned proton decay
searches \cite{prannath, nishino, DUSEL}.

This paper is organized in the following manner. In Sec. II we
present the essence of the model.  Unification of gauge couplings
with left-right intermediate symmetry is examined in Sec.
III where we also discuss proton lifetime predictions.
Derivation of new formulas for the singlet-fermion decay rate and the
CP-asymmetry parameter are  in Sec. IV along with the
predictions for the baryon asymmetry.  In Sec. V we discuss
testable predictions of the model at the Tevatron, LHC and ILC
where we also provide an estimate of the upper  and 
lower bounds on the Drell-Yan pair production cross section.  A
brief summary and conclusions are given in Sec. VI.

\section{The Model}
\label{model}

In this section, we present the salient features of the model
responsible for explaining  neutrino masses, mixings, and
leptogenesis with testable signature at accelerator energies.  We
consider the following pattern of spontaneous symmetry breaking
originating from SUSY $SO(10)$,
\begin{eqnarray}
SO(10) & %\stackrel{54 \oplus 210
\stackrel {(M_U)}
{\longrightarrow} &  SU(2)_L \times
SU(2)_R \times U(1)_{B-L}\times SU(3)_C\times D ~~[{\cal
G}_{2213P}] \nonumber \\ &
\stackrel {(M_P)}
{\longrightarrow} 
& SU(2)_L \times
SU(2)_R \times U(1)_{B-L}\times  SU(3)_C ~~[{\cal G}_{2213}] \nonumber \\
&
\stackrel {(M_R)}
{\longrightarrow}& SU(2)_L \times
U(1)_Y \times  SU(3)_C~~[\std] \nonumber \\
&
\stackrel {(M_Z)}{\longrightarrow}&SU(3)_C \times U(1)_Q \;\;.\nonumber
\end{eqnarray}

\par
The first stage of spontaneous symmetry breaking (SSB) is carried
out by assigning GUT scale vacuum expectation values to the
$\Phi_{54}$ of $SO(10)$ along the direction
singlet\footnote{The r\^ole of this {\em vev} is discussed in
\cite{parida}.} under the Pati-Salam group $ SU(2)_L \times
SU(2)_R\times SU(4)_C \equiv {\cal G}_{PS}$ \cite{ps} as well as
the singlet direction under the left-right gauge group $ SU(2)_L
\times SU(2)_R \times U(1)_{(B-L)}\times  SU(3)_C \equiv {\cal
G}_{2213}$ in the ${\cal G}_{PS}$ multiplet $(1, 1, 15)$
contained in a $\Phi^{(1)}_{210}$ of  $SO(10)$. At
this stage D-parity remains intact and the gauge couplings of
$SU(2)_L$ and $SU(2)_R$ are equal, $g_L=g_R$ \cite{dpar}. The
second stage of SSB takes place by assigning vacuum expectation
value to the D-Parity odd singlet also contained in
$\Phi^{(2)}_{210}$ of $SO(10)$.  By suitable fine tunings of
the trilinear couplings beteen ${\bf 210}$ and the ${\bf {126}_H
\oplus \overline {126}_H}$ or ${\bf {16}_H \oplus \overline
{16}_H}$ the right handed triplets ${\bf \Delta_R \oplus
\overline {\Delta}_R} \subset {\bf {126}_H \oplus \overline
{126}_H}$ and the RH doublets ${\bf \chi_R \oplus \ovl {\chi}_R}
\subset {\bf {16}_H \oplus \ovl {16}_H}$ are made much lighter
compared to their left-handed counterparts. By adopting higher
degree of fine tuning for the RH triplet compared to the RH
doublet, the components of the RH triplet pairs can be assigned
masses between 100 GeV to a few TeV while the RH doublet pairs
are kept heavier, but sufficiently lighter than the GUT scale.
Although we do not ascribe any  {\em vev} directly to the neutral
components of the RH-triplets in ${\bf {126}_H \oplus \overline
{126}_H}$, we will find that once a {\em vev} is assigned to the
neutral component of the RH-doublet in ${\bf {16}_H}$, the
triplet {\em vev} is automatically induced.  Smaller is the
RH-triplet mass fixed by the D-parity breaking mechanism, larger
is the induced triplet {\em vev}.

\par
The reason behind such ordering of Higgs masses and {\em vev}s
becomes transparent once we consider the  Yukawa  Lagrangian near
the intermediate scale emerging from $SO(10)$,      
\ba
{\cal L}_Y = Y \ovl \psi_L \psi_R \Phi + f\psi^T_R\tau_2\psi_R
\bar {\Delta}_R + F\ovl \psi_RT\chi_R + \mu T^TT+H.c.\label{yuklm}
\ea
 where $\psi_{L, R}$ are left- (right-) handed lepton doublets
and $T$ the three fermion singlet fields, one for each
generation.  The superscript $T$, of course, denotes transpose. In
the $(\nu, N, T)$ basis this will lead to a $3\times3$ mass
matrix\footnote{Each entry in this mass matrix is a (3$\times$3)
block.} with vanishing $11$, $13$, and $31$ blocks.
\ba
M_\nu = \pmatrix{\nu & N^c & T}_L \pmatrix{ 0 & m_D & 0  \cr
m_D^T & M_N & M_X\cr 0 & M_X^T & \mu  }
\pmatrix{ \nu \cr N^c\cr T}_L.  \label{matrix}
\ea
Here the $N-T$ mixing matrix arises through the {\em vev} of the
RH-doublet field with $M_X = Fv_{\chi}$, where
$v_{\chi}=\langle \chi^0_R\rangle $, and the RH-Majorana neutrino
mass is generated by the induced {\em vev} of the RH-triplet
with $M_N=fv_R$,  with $v_R=\langle
\overline{\Delta}^0_R\rangle$. The {\em vev} of the weak
bi-doublet $\Phi(2,2,0,1) \subset 10_H$ of $SO(10)$ yields
the Dirac mass matrix for neutrinos, $m_D = Y \langle\Phi^0
\rangle$.

While implementing leptogenesis in this model through $T$ decays,
the out-of equilibrium condition requires the mixing with RH
neutrinos to be small. This  will be naturally obtained if $M_N
\gg M_X $ or if $v_R \gg v_{\chi}$.

Assuming $M_N \gg M_X \gg \mu, m_D$, which would be highly
desirable for the present model, integrating
out the heavy RH-neutrinos leads to the effective
Lagrangian \cite{kang},
\ba
 {\cal  L}_{(mass)} &=&-(\mu - M_X^T M_N^{-1} M_X) ~T^TT -
m_D M_N^{-1} m_D^T ~\nu^T\nu \nonumber\\ && - M_X^T 
M_N^{-1}m_D^T ~{\bar T}\nu + h.c.    \label{eff}
\ea

Interestingly, the  block diagonalization of this mass
matrix results in a cancellation among  the  Type-I see-saw
contributions and  the light neutrino mass $m_{\nu}$  is
dominated by the inverse see-saw and one obtains,
\ba
m_{\nu} &=&~ -m_D~[M_X^{-1}\mu (M_X^T)^{-1}]~m_D^T ,
\label{inv1}
\ea
\vskip -30pt
\ba
   M_T &=& ~\mu -M_X~M_N^{-1}~M_X^T , 
\label{inv2}
\ea
\vskip -30pt
\ba
M &=& ~M_N + ~{M_X}~M_N^{-1}~M_X^T.
\label{inv3}
\ea
It will be shown in the next section that the left-handed
triplets are near the GUT scale while $v_R \sim 10^{10}$ -
$10^{12}$ GeV leading to negligible Type-II contribution  for
light neutrino masses for suitable values of the model
parameters.

\par

To see how the induced {\em vev} is generated, consider the Higgs
superpotential near the intermediate scale where all GUT-scale
masses have decoupled,
\be
W =  M_{\Delta_R}\Delta_R\bar {\Delta}_R + M_{\chi_R}\chi_R\bar {\chi}_R
+\lambda_1 \bar{\Delta}_R\chi_R\chi_R +  \lambda_2 \Delta_R
\bar{\chi}_R\bar{\chi}_R .
\label{supot}
\ee
Using ${ \langle \chi_R^0 \rangle =  \langle {\ovl {\chi}}_R^0
\rangle = v_{\chi}}$, ${ \langle \Delta_R^0
\rangle = \langle {\ovl {\Delta}}_R^0 \rangle = v_{\rm R}}$ which
requires $\lambda_1=\lambda_2 \equiv \lambda$, the vanishing
F-term conditions,  $F_{\Delta^0_R}=F_{\bar
{\Delta}^0_R}=F_{\chi^0_R} =F_{\bar {\chi}^0_R}=0$ give
\be
  { v_R} = - \lambda \frac{v_{\chi}^2}{M_{\Delta_R}}, \;\;
 M_{\Delta_R} M_{\chi_R} = 2\lambda^2{v_{\chi}^2} , \;\;
 M_{\chi_R} = -2\lambda{v_R}.  \label{ivev}  
\ee  
The above equations imply that  even though no direct {\em vev}
is ascribed to ${\Delta_R^0}$ or ${\ovl {\Delta}^0_R}$, a
large  {\em vev} is induced once a direct {\em vev} is
assigned to ${\chi^0_R}$, the latter being essential to
generate the desired $N-T$ mixings.  With lighter RH-triplet
masses $M_{\Delta} \simeq 100$ GeV - 1 TeV,  it is possible to
have $v_R \simeq 10^{10}$ - $10^{12}$ GeV for $v_{\chi} =
10^6$ - $10^7$ GeV.  Since $v_R \gg v_{\chi}$, the spontaneous
breaking $SU(2)_R\times U(1)_{B-L} \to U(1)_Y$ takes place at the
higher scale generating large RH Majorana neutrino masses $M_N
\gg M_X$ leading to small $N_i - T_j$ mixings needed to establish
the out-of equilibrium conditions for leptogenesis.

We assume the Majorana Yukawa coupling to be diagonal, $M_N= {\rm
{diag}} (M_{N_1}, M_{N_2}, M_{N_3})$. This gives $N_i-T_j$ mixing
angles,
\be  
\sin \xi_{ij} \simeq {M_{X_{ij}} \over M_{N_i}}. 
\label{mixxi} 
\ee
In the present model, the left-handed triplet pair in ${\bf
126\oplus \ovl {126}}$ acquire mass near the D-parity breaking
scale  $M_P \gg M_R$. In conventional models even with the
left-handed triplet mass $\simeq 10^{13}$ - $10^{14}$ GeV,
the Type-II see-saw contribution is  comparable to the Type-I
contribution. In this model the Type-II see-saw contribution to
the light neutrino mass matrix is
\ba   
m_{II} = f\lambda^{\prime} \frac{v_{\chi}^2v_u^2}{M_P^2M_{\Delta}}. 
\label{type2}
\ea
where $M_P$ is the D-parity violation scale which is also the
left-handed triplet mass. Now using $M_{\Delta} = 1 $ TeV,
$v_{\chi}= 10^6$ - $10^7$ GeV, $v_u=100$ GeV, and $M_P \simeq
M_U = 10^{16.5}$ GeV, we obtain,
\ba
m_{II} = f\lambda^{\prime} (10^{-20} - 10^{-17}) {\rm ~GeV}, 
\ea
which is at least seven orders of magnitude smaller than the
highest value of hierarchical masses obtained from the neutrino
oscillation data as proposed in \cite{mpr}.

Subject to small RG corrections, the underlying quark-lepton
unification in $SO(10)$ approximates the Dirac neutrino mass
matrix with the up-quark mass matrix.~The light neutrino mass
matrix is constructed using the available data on neutrino masses
and mixings with a reasonable assumption on the leptonic phase of
the PMNS matrix. Our strategy is to determine the mass
eigenvalues and  mixings of fermion singlets as well as their
mixings with RH neutrinos to implement the leptogenesis scenario
through their decays as will be discussed in Sec. \ref{s:lgen}.

Before addressing the leptogenesis issue we show in the next
section that manifest gauge coupling unification occurs in
SUSY $SO(10)$ with  $\lr$ intermediate gauge symmetry. No
nonrenormalizable  ${\rm dim}.5$ operators are needed to support
the unification idea.

\section{Unification, high $W_R$ mass, proton lifetime}

Manifest unification of gauge couplings converging to a
GUT-scale value in SUSY $SO(10)$ models having left-right
intermediate symmetry  has been found possible earlier by
inclusion of additional scalar degrees of freedom beyond those
needed for spontaneous symmetry breaking \cite{lee, mprs,
malinsky}. More recently this method has been evoked to fit
masses of all charged fermions and for explaining  small neutrino
masses with $W_R$-boson mass even at the TeV scale \cite{dev}.
In \cite{mpr} unification of gauge couplings was accomplished by
using threshold-like contributions of  two nonrenormalizable
${\rm dim}.5$ operators at the GUT scale. Manifest unification
has been also found to be possible when both the left-right
intermediate gauge symmetry and SUSY $SO(10)$ are extended to
contain $S_4$ flavor symmetry \cite{parida}. The left-right gauge
symmetry in that case also has unbroken D-parity as well as
unbroken R-Parity down to the intermediate scale.  In the present
model there is no flavor symmetry. D-parity is broken at the
GUT scale and R-Parity is spontaneously broken at a lower scale
by the {\em vev} of RH doublets in $16_H$. In addition the model
has a testable novel feature of accessible doubly charged Higgs
scalars.

In this section we show how manifest unification takes place with
the gauge couplings of ${\cal G}_{2213}$ converging at the
GUT-scale without invoking the effect of any nonrenormalizable
operators. We also show how the proton lifetime for the decay
$p\to e^+\pi^0$ is brought closer to the current experimental
limit \cite{nishino}.

\subsection{Gauge coupling  unification}

We assume the superpartners of the SM particles to have masses of
the order of a TeV.  Using renormalization group equations (RGEs)
for the gauge couplings up to one-loop \cite{gqw}
\ba
{\large {{\mu {{dg_i}\over {d\mu}} =
-\left({{a_i}\over{16\pi^2}}\right)g_i^3}}} ,
\label{rgg}
\ea     
where $i$ ranges over the set of gauge couplings. Below we list
the particles which, with their superpartners, contribute to the
$a_i$ coefficients in different energy ranges.

(i) $\Z \leq \mu \leq M_{SUSY}$:

Here the particle spectrum is the same as in the non-SUSY SM with
three fermion generations,
\ba
a_Y= {{41}\over {10}}, ~~a_{2L} = -{{19}\over {6}}, ~~a_{3C} = -7. 
\label{rgz}  
\ea
(ii) $M_{SUSY} \leq \mu  \leq M_{\sigma}$:

In this range, in addition to the MSSM particle spectrum, we
have the doubly-charged Higgs bosons left  as unabsorbed
components of RH Higgs triplets and these modify only the $a_Y$
coefficients compared to the MSSM.
\ba
a_Y = {{57}\over {5}}, ~~ a_{2L} = 1, ~~ a_{3C} = -3.
\label{rgs}
\ea
Because of  relatively larger value of $a_Y = {{57}\over {5}}$ 
($a_Y ={{33}\over{5}}$ for the MSSM), due to the $\Delta^{\pm\pm}$
near the TeV scale, the $U(1)_Y$ coupling
grows faster, triggering a tendency of unification at
substantially lower scales. This difficulty is bypassed by
embedding $G_{213}$ into the $G_{2213}$ intermediate symmetry. At
the boundary point, the $U(1)_{B-L}$ coupling starts from a lower
value while the $SU(2)_R$ coupling is higher ensuring
 unification at the GUT scale. The exact
unification of all four couplings of $G_{2213}$ is achieved
by introducing additional scalar submultiplets  such as
$\sigma_L(3,0,1)$ and $C_8(1,0,8)$ at scales $M_{\sigma}$ and
$M_C$, respectively. It has been noted earlier that such states
in the adjoint representations of the standard model subgroups
with $Y=0$ could be naturally light  and arise as continuous
moduli states of string theory,  playing a significant role to
reconcile the discrepancy between the GUT scale and the string scale
\cite{bachas}. In our case these submultiplets are contained in
the $SO(10)$ representations $210$ and $45$, whereas $C_8(1,0,8)$
is also contained in the Higgs representation $54 \subset
SO(10)$.  Alternatively, every pair of triplet $\sigma_L$s can
be replaced by a fermion triplet which has been noted to play the
role of stable dark matter \cite{DM} if its mass is low.~This
fermionic state along with others may be present in nonminimal
$SO(10)$ representations \cite{frigham}. 

We will show below that one set of solutions of RGEs needs a
pair of triplet scalars ($n_{\sigma}=2$) or equivalently a
fermionic triplet with mass $\sim 100$ GeV. In that case, only
the scalars $C_8(1,0,8)$ may be treated as naturally light
continuous moduli states of string theory, or, purely from
$SO(10)$ point of view, the mechanism of Refs. \cite{parida, dev}
can  be utilized to make them light by exploiting the generalized
superpotential \cite{fukuyama2}.

Another pertinent question arises if one wishes to use a pair of
moduli states $\sigma_L(3,1,0,1)$ under $\lr$. How is the
lightness of these states ensured in the context of D-Parity
breaking at the GUT scale leading to lighter components of RH
triplets in ${126}_H\oplus {\ovl {126}_H}$ and RH doublets in
${16}_H\oplus {\ovl {16}_H}$.  This question is readily answered
by examining  the part of the superpotential,
\ba
W&=&W_1+W_2+W_3 +....\nonumber \\ W_1&=&M_{126}{\ovl
{126}_H}{126}_H+ {\lambda_{126}}{210}_H {\ovl
{126}_H}{126}_H,\nonumber \\ W_2&=&M_{16}{\ovl
{16}_H}{16}_H+{\lambda_{16}}{210}_H{\ovl {16}_H}{16}_H,\nonumber
\\ W_3&=&M_{45}{45}_H^2+\lambda_{45}{210}_H {45}_H^2 \;\;.
\label{mechnism}
\ea
Noting that the singlet under $\ps$ contained in ${210}_H$ is
D-odd, the RH triplets are made light when the parameters
$M_{126}$ and ${\lambda_{126}} <210_H>$ are in the same phase. Similarly
the RH  doublets are made  lighter than the GUT scale when 
 $M_{16}$ and $\lambda_{16} <210_H>$  are in the same phase.
Thus, it is clear that the same mechanism also renders
$\sigma_L(3,1,0,1)\subset {45}_H$ substantially lighter than the
GUT scale while keeping $\sigma_R(1,3,0,1)\subset {45}_H$ heavy
when  $M_{45}$ and $\lambda_{45}<210_H>$ are in {\em opposite}
phase.

Purely from SUSY $SO(10)$ considerations with standard three fermion
generations, the method of keeping  the relevant Higgs
scalars substantially lighter than the GUT scale has been
discussed in Refs. \cite {parida, dev} by exploiting the
minimization of the generalized superpotential of
Ref.\cite{fukuyama2}.

(iii) $M_{\sigma}  \leq \mu  \leq M_C $: 

In  this range in addition to the contribution of the particles
listed above
 we include those from
$n_{\sigma}$ members of Higgs scalar triplets $\sigma_L (3,0,1)$
leading to  $a_{2L} = 1+2 n_{\sigma}$, 
 and  $a_Y = {{57}\over
{5}}$, ~~ ~~$ a_{3C} = -3$ as before.

(iv) $M_C  \leq \mu  \leq M_R $:
 
Over and above the contributions mentioned above, here we include
the $n_c$ color octets $C_8(1,0,8)$ resulting in ~~$ a_{3C} = -3
+ 3n_C$ and $a_Y = {{57}\over {5}}$, ~~ $a_{2L} = 1+2
n_{\sigma}$, as before.

(v) $M_R  \leq \mu  \leq M_U$:

In the presence of $\lr$ gauge symmetry we have contributions of
all the sub-multiplets discussed above. In addition, from the
${126}_H
\oplus {\overline {126}}_H$ and ${16}_H\oplus {\overline {16}}_H$
the following submultiplets must now be included:\\ $\Phi(2,2,0,
1)\oplus \chi_R (1, 2, -1, 1) \oplus
\ovl {\chi}_R (1, 2, +1, 1)
\oplus \Delta_R (1, 3, -2, 1)
\oplus \ovl {\Delta}_R (1, 3, +2, 1) \oplus n_{\sigma}\sigma_L(3, 1,0, 1)
\oplus n_CC_8(1,1,0,8)$.

(vi) $\mu \ge M_R$ with  $n_{\sigma}= n_C = 3$
 we have
\ba    
a_{BL} = 33/2,~~ a_{2L} =7,~~a_{2R} = 6,~~ a_{3C} = 6 \;\;.
\label{rgr}
\ea
With the above particle content and keeping the possibilities of $M_{\sigma},
M_C$ smaller or larger than the intermediate scale $M_R$, 
allowed solutions are realised with $M_R = 10^9$ - $
10^{12.5}$ GeV, $M_U = 10^{15.75}$ - $10^{16.5}$ GeV and
$\alpha_G^{-1} \simeq 5$ - 10.  This covers the desired range
$M_R = 10^{11}$ - $10^{12}$ GeV required to implement viable 
leptogenesis while satisfying the gravitino constraint.

\begin{center}
\begin{figure}[tbh]
\hskip 4cm
\psfig{figure=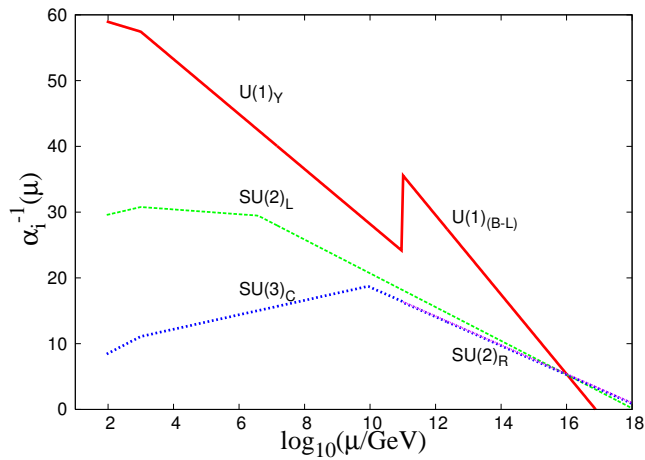,width=8.5cm,height=6.0cm,angle=-90} 
\caption{\em Unification of gauge couplings with left-right
symmetry breaking at $M_R = 10^{11}$ GeV with $n_{\sigma}=n_C=3$
(see text). Below $M_U = M_P  = 10^{16}$ GeV,  $g_L \neq g_R$.
The $SU(3)_C$ coupling  (short-dashed line) and the $SU(2)_R$
coupling (dotted line) appear to merge for $\mu > M_R$ because
fortuitously, in this example, the beta-function coefficients for
both  are nearly equal in this energy range as are the boundary
values of the two couplings at $M_R$.}
\label{evoln}
\end{figure}
\end{center}

For a typical example, the evolution of the  gauge
couplings and unification at the GUT scale are shown in
Fig. \ref {evoln} for which we have obtained
\ba 
M_R = 10^{11} ~{\rm GeV}, ~~ M_U = 10^{16} ~{\rm GeV}, \label{soln}
\ea 
with $\alpha_G^{-1}=5.3$ which is  well within the perturbative
limit.  In  Fig. \ref{evoln} the couplings for $SU(2)_R$ and
$SU(3)_C$ are found to be almost ovelapping above the scale $M_R$
because of a fortuitous identity of their respective beta function
coefficients and near equality of the boundary values at $M_R$ in
this example. The change in slopes at $M_\sigma$ and $M_C$ are
clearly noticeable.
%%%%%%%%%%%%%%%%%%%%%%%%%%%%%%%%%%%%%%%%%%%%%%%%%%%%%%%%%%%%%%%%%%%
%%%%%%%%%%%%%%%%%%%%%%%%%%%%%%%%%%%%%%%%%%%%%%%%%%%%%%%%%%%%%%%%%%%

%%%%%%%%%%%%%%%%%%%%%%%%%%%%%%%%%%%%%%%%%%%%%%%%%%%%%%%%%%%%%%%%%%%%%
\begin{table*}
\begin{center}
\begin{tabular}{|c|c|c|c|}
\hline
$n_{\sigma}$&$n_C$& $M_{\sigma}$&$M_C$\\
         &&(GeV)&(GeV)\\ \hline
$2$&$2$&$100$&$7.16\times 10^6$ \\ \hline
$3$&$3$&$3.8\times 10^6$&$8.67\times 10^9$ \\ \hline
$4$&$3$&$8.63\times 10^8$&$8.67\times 10^9$ \\ \hline
\end{tabular}
\end{center}
\caption{  The number of $\sigma_L(3,0,1)$
and $C_8(1,0,8)$ submultiplets with their respective mass scales
which lead to 
solutions with $M_U=10^{16}$ GeV, $M_R = 10^{11}$
GeV, and $\alpha_G^{-1}=5.3$.}
\label{tabn}
\end{table*}
%%%%%%%%%%%%%%%%%%%%%%%%%%%%%%%%%%%%%%%%%%%%%%%%%%%%%%%%%%%%%%%%%%%%%%%

In Table \ref{tabn} we present several choices of
$n_{\sigma}$ and $n_C$ and their respective  mass scales for which
the same values of $M_R= 10^{11}$ GeV and $M_U = 10^{16}$ GeV 
are obtained as in 
Fig. \ref{evoln}.  The same results follow
when a pair of $\sigma_L$'s are replaced by a fermion triplet
contained in additional $SO(10)$ representations.  These fermions
while driving type-III see-saw for their appropriate mass ranges,
may also serve as stable dark matter candidates if their mass is
low \cite{DM, frigham}.

\subsection{Observable  gauge boson mediated proton decay}

There are elegant methods and models to suppress Higgsino
mediated proton decay  or allow both types of decays  through
dim.5 or dim.6 operators \cite {babubarr, rnm6,babupati2}. In
most of the single-step breaking models, barring a few
\cite{babupati2}, neglecting threshold effects, the unification scale
is $M_U^0 = 2 \times 10^{16}$ GeV with $\alpha_G^{-1}\simeq 25$
which imply large values of the lifetime for gauge boson
mediated proton decay, e.g. $p \to e^+\pi^0$,  for which the
current lower bound is \cite{nishino}  $(\tau_p)_{expt.} \ge
1.01\times 10^{34}$ yrs. Extensive estimations of the decay rate have
been made in minimal GUTs and their extensions with or without SUSY
\cite{prannath, frigham, babupati2, bajc2}.  
Upto a good approximation, the decay width in the present model can be 
written as\\
\ba
\Gamma(p\to e^+\pi^0)&=&\frac{m_p}{64\pi
f_{\pi}^2}\left(\frac{g_{G}^4}{M_U^4}\right)A_L^2{\overline
{\al}_H}^2\left(1+D+F\right)^2\nonumber\\
&&\times\left[(A_{SR}^2+A_{SL}^2)(1+|V_{ud}|^2)^2\right].\label{width}
\ea  
In the above formula ${\ovl {\al}_H}$ is the hadronic matrix element,
$m_p=$ proton mass $=938.3$ MeV, $f_{\pi}=$pion decay constant$=
139$ MeV, and the chiral lagrangian parameters are $D =0.81, F=
0.47$. The short distance renormalization  for relevant ${\rm
dim}.6$ operators evaluated with supersymmetry from $M_U \to
M_{SUSY}$ and without supersymmetry from $M_{SUSY}=1$ TeV $\to
M_Z$ in the present model with appropriate anomalous
dimensions \cite{begn} gives  $A_{SL} \simeq A_{SR} \equiv A_{SD}
= 2.38$. The long distance renormalization factor is known to be
$A_L =1.25$. Noting that $A_R = A_LA_{SD}\simeq 2.98$, and
$F_q=2(1+|V_{ud}|^2)^2 \simeq 7.6$, we then express the lifetime
as

\ba
\Gamma^{-1}(p\to e^+\pi^0)&=& (1.0\times 10^{34} {\rm
yrs.})\left({0.012 ~{\rm GeV}^3}\over
\alpha_H\right)^2
\left({2.98 \over A_R}\right)^2\left({{1/5}\over
\alpha_G}\right)^2\nonumber \\
&&\times\left({7.6\over F_q}\right) \left({M_U \over {1.3\times 10^{16} {\rm
GeV}}}\right)^4 \;\;, \label{lifetime}
\ea

where we have used $\al_H = {\ovl {\alpha}_H}(1+D+F) \simeq
0.012$ GeV$^3$ as per recent lattice estimations
\cite{lattice}.
%######################################################################
%\ba %A_{SD}&=&
%\left(\frac{\alpha_Y(M_Z)}{\alpha_Y(M_S)}\right)^{-\gamma_Y\over
%{2a_Y}} \left
%(\frac{\alpha_Y(M_S)}{\alpha_Y(M_R)}\right)^{-\gamma_Y\over
%{2a'_Y}}\left
%(\frac{\alpha_{2L}(M_Z)}{\alpha_{2L}(M_S)}\right)^{-\gamma_{2L}
%\over %{2a_{2L}}}\left
%(\frac{\alpha_{2L}(M_S)}{\alpha_{2L}(M_{\sigma})}\right)^
%{-\gamma_{2L}\over %{2a'_{2L}}}\nonumber\\ %&& \left
%(\frac{\alpha_{2L}(M_{\sigma})}{\alpha_{2L}(M_U)}\right)^{-\gamma_{2L}%\over
%{2a''_{2L}}}\left
%(\frac{\alpha_{BL}(M_R)}{\alpha_{BL}(M_U)}\right)^{-\gamma_{B%L}\over
%{2a_{BL}}} %\left
%(\frac{\alpha_{2R}(M_R)}{\alpha_{2R}(M_U)}\right)^{-\gamma_{2R}\over
%{2a_%{2R}}}\nonumber\\ %&&\left
%(\frac{\alpha_{3C}(M_Z)}{\alpha_{3C}(M_S)}\right)^{-\gamma_{3C}\over
%{2%a_{3C}}} %\left
%(\frac{\alpha_{3C}(M_S)}{\alpha_{3C}(M_C)}\right)^{-\gamma_{3C}\over
%{2a'%_{3C}}} %\left
%(\frac{\alpha_{3C}(M_C)}{\alpha_{3C}(M_U)}\right)^{-\gamma_{3C}\over
%{ %2a''_{3C}}}.  %\nonumber %\label{sd} %\ea
%#######################################################################
In a number of single-step breaking models or other intermediate
breaking models with $M_U = M_U^0 = 2 \times 10^{16}$ GeV and
$\alpha_G^{-1}\simeq 25$, the one-loop estimation  gives large
proton lifetime $\tau_p(p \to e+\pi^0)\sim O(10^{36})$ yrs. which
is beyond the experimentally  accessible limits of ongoing and
planned proton decay searches  for the $p \to e^+\pi^0$ mode.
 
In the present model some of our predictions  at one-loop level
using eq.(\ref{lifetime}) and RGE solutions are given in Table
\ref{tabtau}.  Although two-loop and threshold corrections are
likely to improve these results, at one-loop level itself our
predictions on the lifetime are substantially less than a large
number of  single step breaking models except few
\cite{babupati2}  and other intermediate breaking models in
conventional SUSY $SO(10)$ GUTs. Our model predictions  are found
to remain within one order of the current experimental limit and
are likely to be accessible to ongoing and planned experiments
for proton decay searches \cite{nishino, DUSEL}. Out of the two
observable model predictions, namely the low mass RH doubly
charged Higgs (discussed later) and proton decay, if any one is
first observed experimentally, the observation on the other
should follow.

%%%%%%%%%%%%%%%%%%%%%%%%%%%%%%%%%%%%%%%%%%%%%%%%%%%%%%%%%%%%%%%%%%%%%%%%%%%%%%
\begin{table*}
\begin{center}
\begin{tabular}{|c|c|c|c|}
\hline
$M_R$&$M_U$& $\alpha_G^{-1}$&$\tau_p(p\to e^+\pi^0)$\\
(GeV)&(GeV) & &(yrs.) \\ \hline
$10^{11}$&$ 1.4\times 10^{16}$&$5.3$&$1.5\times 10^{34}$ \\ \hline
$10^{11}$&$2\times 10^{16}$&$4.2$&$4\times 10^{34}$ \\ \hline
$10^{12}$&$2\times 10^{16}$&$4.1$&$3.8\times 10^{34}$ \\ \hline
$10^{13}$&$2\times 10^{16}$&$3.3$&$2.5\times 10^{34}$ \\ \hline
$10^{9}$&$1.4\times 10^{16}$&$6.2$&$8.6\times 10^{34}$ \\ \hline
\end{tabular}
\end{center}
\caption{Gauge boson mediated decay lifetime for $p\to e^+\pi^0$
in SUSY $SO(10)$ with $G_{2213}$ intermediate symmetry as described in 
the text.} 
\label{tabtau}
\end{table*}
%%%%%%%%%%%%%%%%%%%%%%%%%%%%%%%%%%%%%%%%%%%%%%%%%%%%%%%%%%%%%%%%%%%%%%%%%%%

\section{Leptogenesis  through  singlet fermion decay }\label{s:lgen}

\subsection{Leptogenesis and canonical see-saw}

In the standard formulation of leptogenesis the lightest
right-handed neutrino decays into either $l^-\phi^+$ and
$\nu\phi^0$ or into their conjugate channels $l^+\phi^-$ and
$\bar {\nu}\bar {\phi}^0$ and the desired CP-asymmetry is
generated by the interference of the tree level amplitude with
one-loop amplitudes (vertex and self-energy corrections).
Denoting  the mass eigenvalue of the  $i$th RH neutrino as
$M_{N_i}$, using the canonical Type-I see-saw formula the decay
rate of $N_1$ is, 
\begin{equation}
\Gamma_1 = \frac{1}{8\pi}M_{N_1}(Y_D^{\dagger}Y_D)_{11}
        = \frac{1}{8\pi v_u^2}\tilde {m}_1M_{N_1}^2 \;, \label{gamma1}
\end{equation}
where ${\tilde m}_1$ is roughly the lightest left-handed Majorana
neutrino mass, $v_u$ the {\em vev} of the up-type Higgs, and
$Y_D$ the Dirac-type neutrino Yukawa matrix which, upto RG
corrections, is the same as the up-quark Yukawa matrix.  A net
lepton asymmetry is generated when the decay process goes out of
equilibrium at temperature $\sim M_{N_1}$ satisfying the condition,
\ba
\Gamma_1 &<& H(T=M_{N_1}), \nonumber \\
H(T) &=& 1.66g^{1/2}_*\frac{T^2}{M_{Pl}}, \label{hubble}
\ea
where $H$ is the Hubble expansion rate.  In the normal
hierarchical case  the generated CP-asymmetry is expressed as,
\ba
\epsilon_1=-\frac{3}{8\pi v_u^2}{M_{N_1}\over
M_{N_2}}{Im[(m_D^{\dagger}m_D)_{12}]^2\over{(m_D^{\dagger}m_D)_{11}}}.
\label{estd}
\ea

 The canonical see-saw mechanism gives rise to
the lower bound $M_{N_1}\ge 2.9 \times 10^9$ GeV \cite{di}.
Since this exceeds the upper bound on the gravitino mass by
several orders, the tension between standard leptogenesis and the
gravitino constraint in SUSY theories is explicit.

In  a large class of solutions of the present model, examples of
which are considered in the following subsections, the decay of
two of the three singlet neutrinos to $l^-\phi^+$ and $\nu\phi^0$
(and the charge conjugate states) is kinematically forbidden.
Also, the mass of the remaining singlet neutrino is $\simeq 10^5$
- $10^6$ GeV, determined by the neutrino oscillation data, and is
substantially smaller than that of the RH-neutrinos.  Thus, its
decay can reconcile with the gravitino constraint provided it
goes out-of-equilibrium at temperatures $\sim M_{T_1}$ and it
generates the required lepton asymmetry.  The out-of-equilibrium
condition is found to be naturally achieved due to the small
mixings of $T_1$ with heavy right-handed neutrinos dictated by
the model.

\subsection{Model parameters for the inverse see-saw}

In this  subsection we show how the parameters needed for
leptogenesis are obtained in this model using the inverse see-saw
formula, neutrino oscillation data, up-quark masses and the CKM
matrix elements.  Apart from the RH neutrino mass matrix $M_N$
assumed to be diagonal with the largest element of order $v_R$,
eqs. (\ref{inv1}) - (\ref{inv3}) contain four mass matrices: the
light neutrino mass matrix $m_{\nu}$, the  neutrino Dirac mass
matrix $m_D$, the $N-T$ mixing matrix  $M_X$, and the singlet
fermion mass matrix $\mu$.

We construct $m_{\nu}$ from the mass eigenvalues ($m_{1,2,3}$) via
the PMNS matrix, $U_{PMNS}$, for which we use $\theta_{12}
= 32^\circ$, $\theta_{23}=45^\circ$, $\theta_{13}=7^\circ$ and
take the leptonic Dirac phase $\delta_{PMNS} = 1.0$ radian.
\be
m_{\nu} = U_{PMNS}^T  {\rm ~diag}(m_1, m_2, m_3)U_{PMNS}\;\;.
\label{numass}
\ee
The Dirac neutrino mass matrix is fixed by the underlying
quark-lepton symmetry of $SO(10)$. Neglecting small RG
corrections, it is taken to be approximately equal to the
up-quark mass matrix.  Using the PDG values \cite{pdg} of the CKM
matrix elements, its Dirac phase, and the running masses of the
three up-type quarks, namely, $m_u$ = 2 MeV,  $m_c$ = 1.5 GeV,
$m_t$ = 171 GeV, we have,
\be
m_D\simeq M_U = V_{CKM}^{\dagger}{\rm ~diag}(m_u, m_c, m_t) V_{CKM}, 
\label{dirac} 
\ee
where we have used the CKM phase  $\delta_{CKM} = 1.0$ 
radian, and the quark mixing angles $\sin\theta^q_{12} =
0.2243$, $\sin\theta^q_{23} = 0.0413$, and $\sin\theta^q_{13} =
0.0037$.

The mass matrices $M_X$ and $\mu$ are not constrained by
experimental data. To minimize the unknown parameters we assume a
simple form for the $N-T$ mixing matrix $M_X$,\\
\ba
\bf{M_X} = \left(\br{ccc} M_{X_{11}} & 0 & 0  \\ 0 & 0 & M_{X_{23}} \\0 & 
M_{X_{32}} & 0 \er\right),
\label{mx1}
\ea
where using eqs. (\ref{yuklm}) and (\ref{matrix}) we have defined
\ba
 M_{X_{ij}} = F_{ij}v_{\chi}\;\;. 
\label{mx2}
\ea
With the knowledge of $m_{\nu}$, $m_D$ and   $M_X$ we then use
the inverse see-saw mass formula -- eq. (\ref{inv1}) -- to
obtain elements of the matrix $\mu$ for hierarchical as well as
invertedly hierarchical light neutrino masses.  The $\mu$ and
$M_X$ matrices are used in eq. (\ref{inv2}) to compute the mass
eigenvalues of the singlet fermions and their mixings  by
the diagonalization procedure with,
\ba
\hat {T}_i &=& \sum_j {\tilde U}_{ij} T_j, 
\label{smix}
\ea
where we denote the mass eigenstates by ${\hat {T}_i}$ and the
corresponding mixing matrix by ${\tilde U}$. Thus the two inputs
matrices $M_N$ (chosen diagonal) and $M_X$ -- eq.  (\ref{mx1}) --
completely determine the singlet neutrino, $T_i$, masses and
mixings consistent with the data on the light neutrino mass
spectrum, mixing, and grand unification.

\subsection{Analytic formulas for decay rate and asymmetry parameter}

\begin{center}
\begin{figure}[tbh]
\psfig{figure=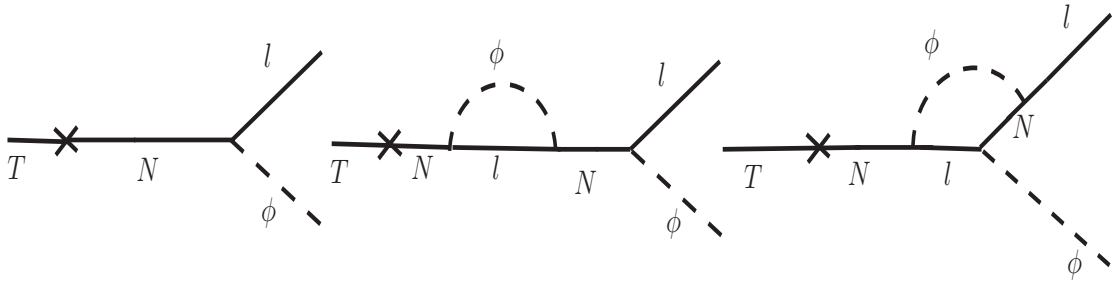,width=15.0cm,height=4cm,angle=0} 
\caption{\em The tree and one-loop level contribution to the
decay of $T_1$ that generate the lepton asymmetry.}
\label{decay}
\end{figure}
\end{center}

The physical processes responsible for leptogenesis are shown in
Fig. \ref{decay} where crosses denote appropriate $N-T$
mixings.  The flavor dependent decay rate for the singlet fermion
$T_i$ of mass $M_{T_i}$ through its mixing with $N_i$ -- recall
eq. (\ref{mixxi}) -- can now be expressed as,
\be
\Gamma_{T_i} = \frac{1}{8\pi v_u^2}M_{T_i}\sum_{jk}
  |\tilde {U}_{ij}|^2\sin^2\xi_{jk}(m_D^{\dagger}m_D)_{kk}.
\label{gammasi}          
\ee
%%%%%%%%%%%%%%%%%%%%%%%%%%%%%%%%%%%%%%%%%%%%%%%%%%%%%%%%%%%%%%%%%%%
%%%%%%%%%%%%%%%%%%%%%%%%%%%%%%%%%%%%%%%%%%%%%%%%%%%%%%%%%%%%%%%%%%%
We find that, depending on the choices of $M_N$ and $M_X$, there
is a wide possibility for the singlet neutrino  mass eigenvalues.
In particular, the model permits a class of solutions where only
one state has mass above the $l\phi$ threshold ($\sim 100$ GeV)
while two others have masses substantially below.  Denoting this
eigenstate as ${\hat {T}}_1$ we discuss leptogenesis through its
decay in the rest of this paper.  Because of the simple assumption
on the $M_X$ matrix given in eq. (\ref{mx1}) the decay rate and
the asymmetry parameter are reduced to the forms,
\ba
\Gamma_{T_1} = \frac{1}{8\pi }M_{T_1}  \frac{K_1}{K_2} & \left[(|{\tilde
U}_{11}|)^2 \sin^2\xi_{11}
(Y_D^{\dagger}Y_D)_{11}
+(|{\tilde U}_{12}|)^2\sin^2\xi_{32}(Y_D^{\dagger}Y_D)_{33}
\right. \nonumber\\
&\left. +(|{\tilde
U}_{13}|)^2\sin^2\xi_{23}(Y_D^{\dagger}Y_D)_{22} \right],
\label{gammas1}          
\ea
where $K_1$,  $K_2$ are modified Bessel functions. 
Even though $Y_D$ is of the same order as the up-quark
Yukawa matrix, the smallness of $\Gamma_{T_1}$ compared to the Type-I
see-saw case --  eq. (\ref{gamma1}) -- originates from two sources:
(i) Allowed values of $M_{T_1}\ll M_{N_i} (i=1,2,3)$, (ii)
$\sin^2{\xi}_{jk} \ll 1 ~(j, k = 1, 2, 3)$. These two features achieve
the out-of-equilibrium condition at temperature $\sim M_{T_1}$
satisfying the gravitino constraint.  The asymmetry parameter can be
expressed as,
\ba
 \epsilon_1 &=& \frac{-3 M_{T_i}}{8\pi}\frac{P}{Q}, \nonumber\\
{P} &=&  \left [(|{\tilde U}_{11}|)^2 \sin^2 \xi_{11}/M_{N_3}
- (|{\tilde U}_{12}|)^2 \sin^2 \xi_{32}/M_{N_1}\right]
  Im[Y_{1i}Y_{3i}^*]^2 \nonumber\\
&&+ \left[(|{\tilde U}_{13}|)^2 \sin^2 \xi_{23}/M_{N_3}
- (|{\tilde U}_{12}|)^2 \sin^2 \xi_{32}/M_{N_2}\right ]
Im[Y_{2i}Y_{3i}^*]^2 \nonumber\\
&&+ \left[(|{\tilde U}_{11}|)^2 \sin^2 \xi_{11}/M_{N_2}
- (|{\tilde U}_{13}|)^2 \sin^2 \xi_{23}/M_{N_1}\right ]
Im[Y_{1i}Y_{2i}^*]^2, \nonumber\\
{Q} &=& {~|{\tilde {U}_{11}}|}^2\sin^2\xi_{11}(Y_D^{\dagger}Y_D)_{11}
+{|{\tilde {U}_{12}}|}^2\sin^2\xi_{32}(Y_D^{\dagger}Y_D)_{33}\nonumber\\
&&+ |{\tilde U}_{13}|^2\sin^2\xi_{23}(Y_D^{\dagger}Y_D)_{22} \;\;.
\label{epsilon}
\ea
In Table \ref{t:numass} we present for two typical solutions 
the right-handed Majorana neutrino masses, matrix
elements of $M_X$, the mixing matrix ${\tilde U}$,
and the $T$-particle masses when the light neutrino masses are
hierarchical or invertedly hierarchical. $\epsilon_1$ is obtained
through eq. (\ref{epsilon}).
%%%%%%%%%%%%%%%%%%%%%%%%%%%%%%%%%%%%%%%%%%%%%%%%%%%%%%%%%%%%%%%%%%%%%%%%%%%%%%
\begin{table*}
\begin{center}
\begin{tabular}{|c|c|c|c|c|c|c|c|c|c|c|}
\hline
Hier.&$M_{X_{11}}$& $M_{X_{23}}$&$M_{X_{32}}$
&$M_{T_{i}}$&$m_i$&$\tilde{U}_{11}$& $\tilde{U}_{12}$&
$\tilde{U}_{13}$&$\epsilon_1$  \\ 
 & $(10^5 {\rm GeV})$ & $(10^5 {\rm GeV})$& $(10^5 {\rm GeV})$&
$({\rm GeV})$& $(10^{-2} {\rm eV})$& & & & $(10^{-7})$\\ \hline &     &
&    &1.11$\times 10^5$&1.1 &   &  &   & \\ NH & 1.2 & 0.18 & 18
&92 & 1.4 & -0.992 & -0.125 & 0.035 & 1.11 \\ &
&      &    &7.7&5.2 &   &  &   & \\ \hline &     &      &
&1.83$\times 10^6$&5.3 &   &  &   & \\ IH & 1.2 & 60.0 & 1.8 &65&
5.4 & -0.085 & -0.001 & 0.996 & 1.40 \\ &     &
&    &9&2.0 &   &  &   & \\ \hline
\end{tabular} 
\end{center} 
\caption{Sample solutions with one singlet, $T_1$, at the
right mass scale for leptogenesis. For both normal (NH)
and inverted (IH) hierarchies the masses of the singlet neutrinos
and the light neutrinos are displayed. $M_{N} = {\rm
{diag}}(2\times 10^{7},  5\times 10^{10}, 9\times 10^{10})$ GeV
has been chosen for both cases. $\kappa =
0.40 -0.55$ for agreement with the observed baryon asymmetry.}
\label{t:numass}
\end{table*}

\subsection{The baryon asymmetry}

For large departure from equilibrium in the $T_1$ decay, the
lepton asymmetry per unit entropy at temperature $T > M_{T_1}$ is
\cite{covi}
\begin{equation}
 {n_L\over s } \simeq \frac{\kappa \epsilon_1}{s}\frac
{g_{T_1}T^3}{\pi^2} =
\frac{45}{2\pi^4}\frac{g_{T_1}}{g_*}\kappa\epsilon_1 = 4.33 \times
10^{-3}\kappa \epsilon_1,
\label{lasy}
\end{equation}
where $\kappa$ is the efficiency factor and $g_{T_1} = 2$ the
number of degrees of freedom of $T_1$. The entropy density $s =
(2/45)g_*\pi^2 T^3$ where $g_*= 106.75$, the effective number of
relativistic degrees of freedom contributing to entropy in the
standard model.  Denoting by $N_H$ the number of Higgs doublets
($N_H$=1 in this model), the baryon to entropy ratio is:
\begin{equation}
{n_B\over s} = -\frac {24+4N_H}{66+13N_H}{n_L\over s }
             = -\frac{28}{79}  {n_L\over s }  
             = -1.53\times 10^{-3}\kappa\epsilon.
\label{bstols}
\end{equation}
Noting that $s = 7.04n_{\gamma}$, where $n_{\gamma}$ is the
photon density, the observed baryon asymmetry is,
\begin{equation}
\eta_B \equiv  {n_B\over n_{\gamma}}   \simeq 10^{-2}\kappa\epsilon_1.
\label{bauexp}
\end{equation}
This is to be compared with  \cite{wmap}:
\be
(\eta_B)_{\rm expt} = (6.15 \pm 0.25)\times 10^{-10}.
\label{baudat}  
\ee
We find that  for both the cases (NH as well as IH) the
predictions are in  agreement with the observed value. In Table
\ref{t:numass} we have exhibited only two out of a large number
of allowed solutions with efficiency factors $\kappa = 0.4 -
0.5$.

For comparison, in Refs. \cite{covi, frigerio} maximal
efficiency, $\kappa \simeq 1$, has been considered to obtain the
requisite baryon asymmetry. In Ref. \cite{frigerio} constraints
on the Dirac Yukawa coupling of the RH neutrino has been examined
and it turns out to be small. In our model the effective Yukawa
coupling of the decaying particle $T_1$ (instead of $N_1$) to the
$l\phi$ pair is  essentially modified by the product of two extra
factors each of which is a mixing substantially smaller than
unity. Thus,  the effective Yukawa coupling of the
decaying singlet neutrino remains small.

The present model permits a variety of solutions with $\epsilon_1
\simeq 10^{-6}$ - $10^{-8}$ which match the observed
baryon asymmetry when the efficiency factors $\kappa \simeq
{\cal O}(10^{-2}) - {\cal O}(1)$.
 
\section{Right-handed doubly charged Higgs at colliders}

In this section we  briefly discuss how this model can be
experimentally tested at high energy colliders such as the
Tevatron,  LHC or  ILC. The light doubly-charged Higgs boson
provides a clear scope for this. We relate the production cross
section of the $\Delta_R^{\pm\pm}$ with that of a $\Delta_L^{\pm\pm}$ of
the same mass\footnote{In our model the $\Delta_L^{\pm\pm}$ is very
heavy due to D-Parity breaking.}.  This relationship permits the
setting of upper and  lower bounds on the pair production cross
sections at the Tevatron or LHC in a model-independent manner.

As explained in Sec. \ref{model}, the RH-triplets in ${\bf
{126}_H\oplus {\overline {126}}_H}$ carrying $B-L= \pm 2 $ are
made light in this model through the D-parity breaking mechanism
with the component  masses $M_{\Delta_R} \simeq 100$ GeV to a few
TeV.  This enhances the induced {\em vev}, $v_R$, resulting in
high scale LR gauge symmetry breaking with large $W_R^{\pm}$, $
Z_R$ gauge boson masses and heavy RH Majorana neutrinos.  The
largeness of the  RH Majorana masses in the model lead to
naturally small $T$-$N$ mixings essential to satisfy the
out-of-equilibrium condition for the $T$-decay to drive
leptogenesis.

Because of large masses of the RH gauge bosons, $m_{W_R^{\pm}},
m_{Z_R} \sim M_R = 10^9$ - $10^{15}$ GeV, the decays
$\Delta_R^{\pm\pm} \to W_R^{\pm} W_R^{\pm}$ are kinematically
forbidden and so only leptonic decays are possible. The $SO(10)$
invariant Yukawa interaction with fermions given in eq.
(\ref{yuklm}), $f_{ij}.16_i. 16_j. {\ovl {126}_H}$, makes the
doubly charged Higgs boson leptophilic, its only  decay
modes being $\Delta_R^{\pm\pm} \to l_{R_i}^{\pm}l_{R_j}^{\pm}
(i,j=1,2,3)$.

Due to their heaviness, $W_R$ and $Z_R$ also play no r\^{o}le
in $\Delta^{\pm\pm}$ production at the Tevatron or LHC.   The
production will dominantly be through the electromagnetic or
$Z^*$-exchange Drell-Yan mechanism.

The couplings $f_{ij}$ determine the branching ratios of the
$\Delta_R^{\pm\pm}$ to the different leptonic final states.
Although for the sake of economy and simplicity we have chosen
$f_{ij}$ to be diagonal in the previous section, our choice has been
guided by negligibly small values of the non-diagonal elements 
suggested by current limits on lepton flavor violating decays
such as $\mu \to 3e$ and $\tau \to 3e$. Out of a large number of
possible solutions, the two sets  given in Table \ref{t:numass}
have $M_{N_1}=2\times 10^{7}$~GeV, ~~$M_{N_2}=5\times
10^{10}$~GeV,~~$M_{N_3}=9\times 10^{10}$~GeV, which for $v_R
\simeq  10^{11}$ GeV corresponds to $f_1 \equiv f_{ee} = 0.0002,
f_2 \equiv f_{\mu\mu} = 0.5, f_3 \equiv f_{\tau\tau} = 0.9$.
Keeping in mind the wide classes of solutions permitted in this model,
we will discuss possible implications for $f_1$ = 0.0002 - 0.001, and
$f_2 \leq f_3$.

The mass ordering of the $W_R^\pm$, $Z_R$, and
$\Delta_R^{\pm\pm}$ discussed above is specific to this model.
Interestingly, low mass doubly charged Higgs bosons with similar
interactions have been shown to be generic in a class of SUSYLR
models with $ M_{W_R} \ge 10^9 $ GeV  which require
non-renormalizable terms in the superpotential \cite{rnm4,
chacko, dutta}. Within  the non-SUSY left-right model  prospects
of Drell-Yan pair production of LH doubly charged Higgs bosons at
high energy colliders and their detection \cite{arc} and the
impact of QCD corrections thereon  \cite{spira} have also been
investigated.

\subsection{Bounds from muonium-antimuonium conversion }

Muonium($M$)-Anti-muonium($\bar M$) conversion, $\mu^+e^- \to
\mu^-e^+$ can be mediated by $\Delta_R^{\pm\pm}$ giving rise to
an effective coupling \cite{rnm5},
\ba
G_{M-{\bar M}} \simeq \frac{f_1f_2}{4\sqrt{2}M_{\Delta}^2}.\label{mm}
\ea
Experimental searches for this transition 
yields the upper bound \cite{willmann}
\be
G_{M-{\bar M}} \le 3\times 10^{-3} G_F, \label{mmb}
\ee

where $G_F=1.17 \times 10^{-5}$ GeV$^{-2}$ is the Fermi coupling.  
Combining  eqs. (\ref{mm}) and (\ref{mmb}) gives 
\ba
M_{\Delta} \ge \left (\frac{f_1f_2}{12 \times 10^{-3} \sqrt {2}
G_F}\right)^{1/2}.\label{deltab}
\ea

For example, choosing $f_1\sim f_2 \sim 1$, a condition
applicable to quasi-degenerate RH neutrinos but which is outside
the presently allowed solutions, we obtain $M_{\Delta} \ge$
$2.24$ TeV which is beyond the Tevatron limit, but within the LHC
range.  But when we use the class of solutions which permit $f_1
\simeq 0.0002$ and $f_2 \simeq 0.5$, we obtain  $M_{\Delta} \ge
22.4$ GeV. This is not inconsistent with the
experimental search limit reached by DO and CDF collaborations at
Fermilab Tevatron with $M_{\Delta} \ge 112$ GeV/c$^2$ and $M_{\Delta}
\ge 127$ GeV/c$^2$, respectively \cite {CDF, DO}.\\

\subsection{Drell-Yan pair production at LHC or Tevatron}\label{ss:DY}

Previous searches at LEP have already excluded
$\Delta_R^{\pm\pm}$ below $97$ GeV/c$^2$ \cite{LEP}.  At hadron
colliders the doubly-charged Higgs boson will be dominantly
created through pair production via the basic Drell-Yan process
$q{\bar q}\to \gamma^*, Z^* \to \Delta_i^{++}\Delta_i^{--},(i=L,
R)$.  At the quark level, the cross section depends only on the
quantum numbers and mass of the doubly charged scalars.  In the
present model with supersymmetry and purely renormalizable
interactions, the only allowed decay mode  is,
\ba
 \Delta_R^{++} \to l_R^+l_R^+
\ea 
and its conjugate.
At the Fermilab Tevatron or the LHC, the production of the
doubly charged boson will be through
\ba
 {\rm p} {\bar {\rm p}} ({\rm p}) \to (\gamma^*, Z^*) X \to
\Delta_R^{++}\Delta_R^{--} X\to l_R^+ l_R^+l_R^{'-}l_R^{'-} X
\ea 
where for the dominant modes in our model $l,l^{'} = \tau, \mu$
since $f_1 \ll f_2 \simeq f_3 \simeq 1$. 

The parton level $\Delta_i^{\pm\pm}$
pair production cross section through $\gamma^*$ and $Z^*$ exchange
is expressed as \cite{dutta,spira}
\ba
{\hat {\sigma}_{i}} &=& \frac{\pi\alpha^2\beta_i^3}{9{\hat
{s}}}\Sigma_i^q,  \nonumber\\
\Sigma_i^q &=&
\left[ Q_q^2Q_{\Delta_i}^2+\frac{{\hat
{s}}^2(g_{qA}^2+g_{qv}^2)g_{\Delta_i v}^2}{({\hat
{s}}-M_Z^2)^2+\Gamma_Z^2M_Z^2} \right. \nonumber \\ 
 &&\left. +\frac{2{\hat
{s}}Q_qQ_{\Delta_i}({\hat {s}}-M_Z^2)g_{qv}g_{\Delta_i v}}{({\hat
{s}}-M_Z^2)^2+\Gamma_Z^2M_Z^2}\right],~~i=L, R. \label{rx}
\ea
\par\noindent 
Here ${\hat s} \equiv Q^2 = \tau s$ is the square of the c.m.
energy of the colliding quark-antiquark pair and $\tau$ the
product of momentum fractions carried by them.  $\alpha$ is the
fine-structure constant at the relevant energy scale and $\beta_i =
\sqrt{(1- 4M_{\Delta_i}^2/{\hat s})}$  is the velocity of the
doubly charged Higgs, $\Delta_i^{\pm \pm}$, produced in the
collision ($i= L, R$).  $g_{qv}=
(I_{3q}-2Q_q\sin^2\theta_W)/(2\sin\theta_W\cos\theta_W),
g_{qA}=I_{3q}/(2\sin\theta_W\cos\theta_W)$ for the quark $q$ and 
$g_{\Delta_i v}= (I_{3\Delta_i}-Q_{\Delta_i}s_W^2)/(2s_wc_w)$.
$Q_q(Q_{\Delta_i})$ is the electric charge number of the quark $q$ (Higgs
$\Delta_i$) and  $I_{3q}(I_{3\Delta_i})$ the third component of
$SU(2)_L$ isospin for $q$ ($\Delta_i$).

It is clear from eq. (\ref{rx}) that for 
\be
\hat s > 4M_{\Delta_i}^2 \gg M_Z^2,\label{limit}
\ee
the $\Sigma_i^q$ become independent of ${\hat s}$ and
the momentum fractions carried by the quarks leading to 
\be
\Sigma^q_i \to {\overline {\Sigma}_i^q}
=\left[ Q_q^2Q_{\Delta_i}^2+(g_{qA}^2+g_{qv}^2)g_{\Delta_i v}^2  
+2Q_qQ_{\Delta_i}g_{qv}g_{\Delta_i v}\right], (i= L,R).\label{limitv}
\ee

Noting that $I_{3\Delta_i} = 1 (0)$ for $i= L(R)$ we obtain from
eq. (\ref{limitv})
\be
{\overline {\Sigma}_R^u}= 0.59 {\overline {\Sigma}_L^u}, \;\;\;
{\overline {\Sigma}_R^d}= 0.79 {\overline {\Sigma}_L^d}.\label{rlations}
\ee
These relations translate to  upper and lower bounds on the
production cross section of the $\Delta_R^{\pm\pm}$ which do not
depend on the proton structure functions or the model-origin of
these Higgs bosons as long as the gauge symmetry at the
electroweak scale is the standard model. Thus,
\be
0.59\sigma_L \le \sigma_R \le 0.79\sigma_L. \label{bounds}
\ee   
The above bounds for Drell-Yan pair production relate the cross
section for RH doubly charged Higgs at LHC or Tevatron with that
for their LH counterparts with the same mass provided the mass
is $\ge$ 150 GeV.

Using these relations and the published results for $\Delta_L^{\pm\pm}$
production one can estimate the number of events in this model.
For example\footnote{We thank Anindya Datta for providing these
numbers.}, at the LHC, for $m_{\Delta_L^{\pm\pm}}$ = 200 GeV (1
TeV) the Drell-Yan production cross section in fb is
49.4, 96.4, 169.5 (0.004, 0.04, 0.14) for $\sqrt{s}$ = 7, 10, or
14 TeV, respectively. Using eq. (\ref{bounds}) and assuming
an integrated luminosity of 30 fb$^{-1}$ for
$m_{\Delta_R^{\pm\pm}}$ = 200 GeV one would expect 1022, 1995,
3509 events for the three cases. If $m_{\Delta_L^{\pm\pm}}$ = 1
TeV then for $\sqrt{s}$ = 10 TeV one would require about 200
fb$^{-1}$ integrated luminosity for a 5-event signal.

  At the Tevatron with ${\sqrt s} = 2$ TeV, $\sigma_R \simeq 12-16$
fb for $M_{\Delta_R} = 150 $ GeV and with an integrated
luminosity of $350$ pb$^{-1}$ the predicted number of events is
nearly $4-6$. With an acquired integrated luminosity of $10$
fb$^{-1}$, the mass reach of up to $M_{\Delta_R} = 300$ GeV can be
achieved where $\sigma_R \simeq (5-7)\times 10^{-2}$ fb.

 In our model since $f_1 \ll f_2 \simeq f_3$ , the dominant decay
modes of the produced pair would be through the following four
lepton channels every one of which would be almost equally
likely\\ $\Delta_R^{++}\Delta_R^{--} \to
\tau_R^+\tau_R^+\tau_R^-\tau_R^-,\mu_R^+\mu_R^+
\mu_R^-\mu_R^-, \tau_R^+\tau_R^+\mu_R^-\mu_R^-,
\tau_R^-\tau_R^-\mu_R^+\mu_R^+$.

The standard model backgrounds for such production processes have
been discussed in \cite{arc}. The signal event should have
negligible missing $p_T$. Moreover, the two pairs of like-charged
leptons are constrained to each have the invariant mass equal to
$m_{\Delta_R}$. These criteria and a judicious cut on the
$l^+l^-$ pair invariant mass to remove $ZZ$ contributions
effectively removes the entire background. 

Unlike the Tevatron and the LHC where the doubly charged bosons
are pair-produced, at proposed muon colliders  resonant
production of these bosons  could take place if $\mu^-\mu^-$
colliders are arranged. The singly produced $\Delta^{--}$ would
decay via $2\mu$ or $ 2\tau$ channels providing the cleanest
signals for these bosons.  In contrast to a large
class of asymmetric left-right models where the decay of the
right-handed doubly charged bosons could proceed via
kinematically allowed channels such as $\Delta_R^{++}\to W_R^+W_R^+,
W_R^+\Delta_R^+,\Delta_R^+\Delta_R^+$, this model allows decay
only in the bilepton channel providing a signature of its
genuine leptophilic property.

\section{Summary and conclusions}

In summary, we have implemented flavor-dependent leptogenesis
through the decay of singlet fermions with masses $M_{T_1} =10^5$
- $10^6$ GeV in SUSY $SO(10)$ while satisfying the gravitino
constraint.  The left-right intermediate symmetry is at a high
scale corresponding to $W_R$ and $Z_R$ masses larger than
$10^{11}$ GeV.  This has been made possible by using the RH
triplets in ${\bf {126}_H\oplus {\overline {126}}_H}$ as well as
the RH doublets in ${\bf {16}_H\oplus {\overline {16}}_H}$. Not
only is the singlet fermion-RH-neutrino mixing generated by the
{\em vev} of ${\bf {\overline {16}}_H}$ but also this mixing
becomes naturally small through the large {\em vev} of the
RH-triplet induced by the doublet {\em vev}. In addition to
obtaining renormalizable mass for the RH neutrino through this
mechanism, manifest unification of gauge couplings is also
achieved purely by renormalizable interactions and, thus,
non-renormalizable dim.5 operators used earlier are dispensed
with. In contrast to the earlier attempts where an assumed value
of the CP-asymmetry parameter was shown to yield the lepton
asymmetry numerically, in this work we have derived and suggested
new analytic formulas leading to the correct asymmetry parameter
and the observed baryon to photon density ratio. We have found
that both the decay rate and the CP-asymmetry are explicitly
flavor dependent.  Whereas our previous work \cite{mpr} required
a normal  hierarchy, here we have also found successful
implementation  in the case of inverted hierarchical
neutrino masses. Unlike a host of low-scale leptogenesis models,
this model works with hierarchical heavy RH-neutrino masses and
no resonant condition with extreme degeneracy among them is
needed. A decisive test of the present model would be through the
detection of doubly charged Higgs bosons $\Delta_R^{\pm\pm} \to
l_R^{\pm}l_R^ {\pm}$ at the Tevatron, LHC, or a future muon collider.
The model provides an example of truly leptophilic doubly charged
Higgs bosons.  Without using parton-density distribution
functions we have also shown in a model independent manner that
the pair production cross sections for RH doubly charged Higgs at
Tevatron or LHC energies are bounded between $59\%- 79\%$ of
their LH counterparts with same masses.

As the unification, triggerred by low-mass doubly charged bosons,
occurs with a large (but perturbative) unified gauge coupling, the
decay lifetime $\tau_p(p\to e^+\pi^0)$ is substantially reduced
compared to conventional SUSY GUTs and remains within one order
of the current experimental limit.  This is likely to be
accessible to the ongoing and planned proton deacy searches.
The model appears to be rich in dark matter candidates which will be
investigated elsewhere.\\  

\par\noindent{\bf ACKNOWLEDGMENTS}\\

The authors acknowledge  useful discussions with Professor R. N.
Mohapatra. They also thank Dr. Anindya Datta for help with some of the
material in section \ref{ss:DY}. M.K.P.
thanks the Harish-Chandra Research Institute (HRI), Allahabad for
hospitality. The work of A.R. is supported by the XI Plan
`Neutrino Physics' and RECAPP projects at HRI.\\

\end{document}